\documentclass[anonymous=false,acmsmall,screen,nonacm]{acmart}
\usepackage{balance}
\usepackage{subcaption}
\usepackage{xcolor}
\usepackage{soul}

\AtBeginDocument{%
  \providecommand\BibTeX{{%
    \normalfont B\kern-0.5em{\scshape i\kern-0.25em b}\kern-0.8em\TeX}}}

\usepackage{enumitem} %better  enumerate, itemize, and description
\usepackage{longtable}

\begin{document}


\title[Portobello: Extending Driving Simulation from the Lab to the Road]{\textsc{Portobello}: Extending Driving Simulation \\from the Lab to the Road}

\author{Fanjun Bu}
\affiliation{%
  \institution{Cornell Tech}
  \streetaddress{2 W Loop Road}
  \city{New York}
  \country{U.S.}}
\email{fb266@cornell.edu}

\author{Stacey Li}
\affiliation{%
  \institution{Cornell Tech}
  \streetaddress{2 W Loop Road}
  \city{New York}
  \country{U.S.}}
\email{sl3326@cornell.edu}

\author{David Goedicke}
\affiliation{%
  \institution{Cornell Tech}
  \streetaddress{2 W Loop Road}
  \city{New York}
  \country{U.S.}}
\email{dg536@cornell.edu}

\author{Mark Colley}
\email{mark.colley@uni-ulm.de}
\affiliation{%
  \institution{Ulm University}
  \city{Ulm}
  \country{Germany}
}
\affiliation{%
  \institution{Cornell Tech}
  \streetaddress{2 W Loop Road}
  \city{New York}
  \country{U.S.}
}

\author{Gyanendra Sharma}
\affiliation{%
  \institution{Woven by Toyota}
  \city{Los Altos}
  \state{California}
  \country{USA}
}
\email{gyanendra.sharma@woven.toyota}

\author{Hiroshi Yasuda}
\affiliation{%
  \institution{Toyota Research Institute}
  \city{Los Altos}
  \state{California}
  \country{USA}
}
\email{hiroshi.yasuda@tri.global}

\author{Wendy Ju}
\affiliation{%
  \institution{Cornell Tech}
  \streetaddress{2 W Loop Road}
  \city{New York}
  \country{U.S.}}
\email{wendyju@cornell.edu}

\renewcommand{\shortauthors}{Bu, et al.}

\begin{abstract} 
In automotive user interface design, testing often starts with lab-based driving simulators and migrates toward on-road studies to mitigate risks. Mixed reality (XR) helps translate virtual study designs to the real road to increase ecological validity. However, researchers rarely run the same study in both in-lab and on-road simulators due to the challenges of replicating studies in both physical and virtual worlds. To provide a common infrastructure to port in-lab study designs on-road, we built a platform-portable infrastructure, Portobello, to enable us to run twinned physical-virtual studies. As a proof-of-concept, we extended the on-road simulator XR-OOM with Portobello. We ran a within-subjects, autonomous-vehicle crosswalk cooperation study (\textit{N}=32) both in-lab and on-road to investigate study design portability and platform-driven influences on study outcomes. To our knowledge, this is the first system that enables the twinning of studies originally designed for in-lab simulators to be carried out in an on-road platform.
\end{abstract}

\begin{teaserfigure}
  \includegraphics[width=\textwidth]{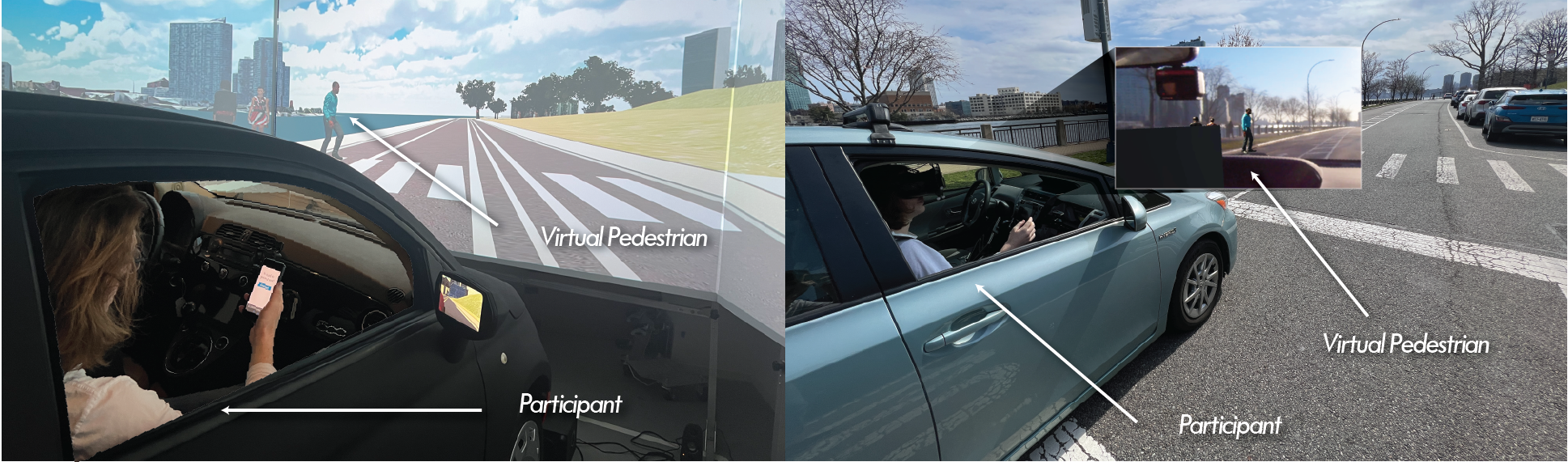}
  \caption{Portobello enables \textit{platform portability}, so that the same study can be run on in-lab (left) and on-road (right) driving simulation platforms.}
  \Description{The picture is split into two halves. In the left half, a black Fiat 500 stands in front of screens on which a driving scene is displayed. The driving scene shows a pedestrian crossing on a crosswalk. In the Fiat 500, there is a participant holding a smartphone in the passenger's seat. In the right half, a Toyota Prius stands at the real crosswalk. Again, a participant sits in the passenger's seat and wears a Varjo XR 1 video-passthrough headset. No real pedestrian is visible but the headset shows a virtual pedestrian crossing.}
  \label{fig:teaser}
\end{teaserfigure}

\maketitle
\begin{figure*}[ht]
    \centering
    \includegraphics[width=\textwidth]{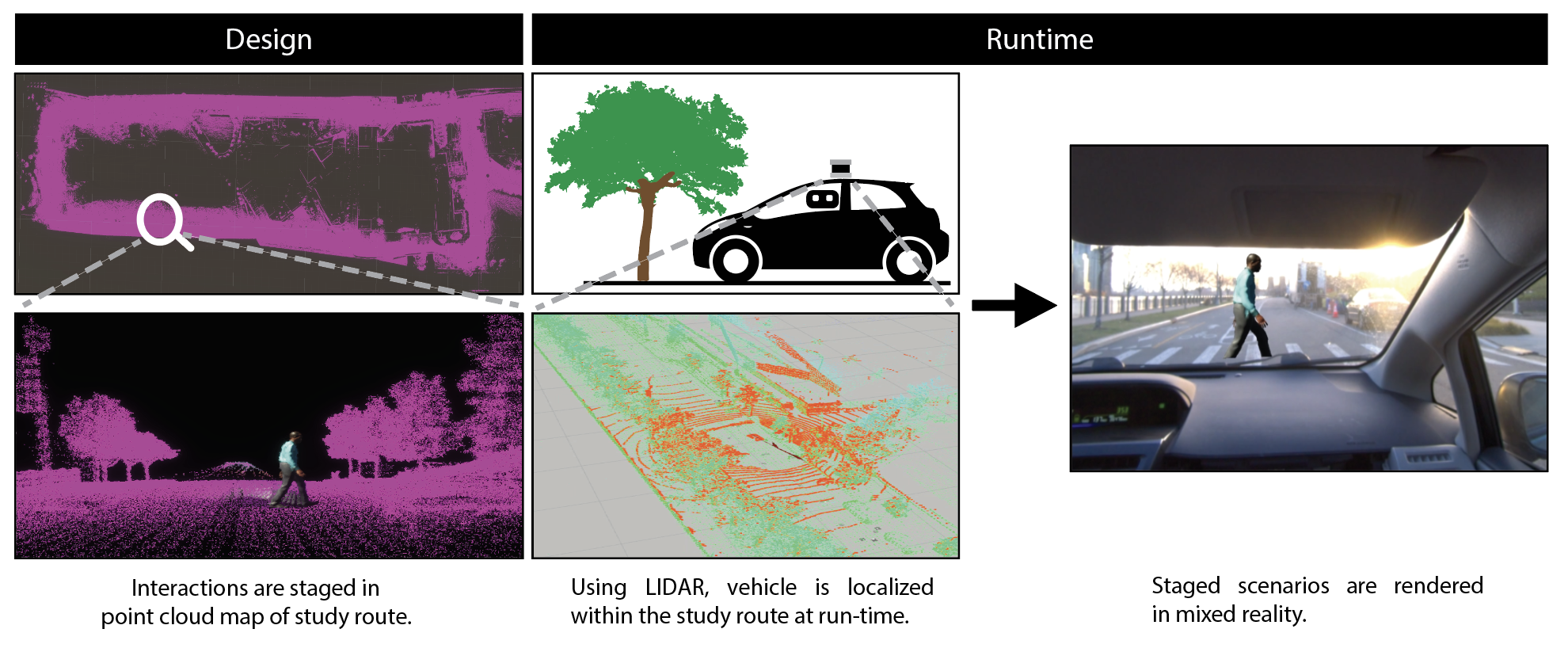}
    \caption{Complete execution pipeline using the Portobello infrastructure. During the design phase, researchers can drag and drop virtual objects on the point cloud map. At run time, the LiDAR-based navigation system locates the vehicle's position within the same map and sends the relative vehicle position to Unity. As a result, passengers wearing the video-see-through headset can see virtual objects at their corresponding real-world locations.}
    \label{fig:xroom_system}
    \Description{This figure shows the execution pipeline using the Portobello infrastructure. It is divided into two phases: the design phase and the runtime phase. Interactions are staged in the point cloud map of the study route. The map has to be recorded prior and is visualized in purple. During runtime, the application uses LiDAR to localize itself within the same map, and stage scenarios are rendered in mixed reality based on the vehicle's real-time location. The far-right figure shows the scene of the virtual pedestrian crossing the real crosswalk. Note that the point cloud map is only rendered in Unity in the design phase. At runtime, it is only loaded into ROS localization algorithms for tracking purposes. It will not be rendered in Unity for mixed-reality display.}
\end{figure*}

\section{Introduction}
Driving simulators play a critical role in human-centered automotive research
applications because they allow people to experience different driving
scenarios in a safe and repeatable fashion. Researchers have full control over
the simulation setup and can program complex events in virtual environments.
However, one of the major challenges for driving simulation has always been the
difficulty of replicating the inertial forces and vehicle dynamics present in
on-road driving \cite{pinto2008development}--even the most high-end motion platform driving simulators only
replicate a fraction of the forces felt in real-world
driving~\cite{greenberg2003effect}. These forces are more critical in testing
automated vehicle (AV) scenarios, where participants are often given
non-driving-related tasks that keep them from looking at the screens of the
simulations (e.g., \cite{10.1145/3342197.3344531}). Without the physical
sensation and the visual engagement of the simulation environment, the
immersiveness of the environment drops, making the evaluation of novel user
experience and interaction techniques such as VEmotion~\cite{VEmotion2021} or
SoundsRide~\cite{SoundsRide2021} difficult. One way to address this problem is
to incorporate driving simulation into a real vehicle driving on real streets
(on-road simulators)~\cite{baltodano2015rrads}.

On-road simulators are possible due to the maturity of XR, where digital
displays blend reality and virtuality to increase the level of a user's
immersion. The use of XR-in-the-car using simulated or actual dash-mounted
heads-up displays have been explored by prior researchers such as
\citet{tonnis2005experimental}, \citet{kim2009simulated},
\citet{schall2013augmented}, \citet{ARCAR2020}, \citet{PassengXR2022},
\citet{10.1145/3411764.3445351}, \citet{von2019increasing},
\citet{narzt2006augmented}, and \citet{bark2014personal}.

Despite this influx of on-road simulators under development, it remains
challenging to migrate studies from in-lab simulators to on-road simulators.
From a study design perspective, the key difference between in-lab and on-road
simulators is that with in-lab simulators, designers have full access not only
to the virtual vehicle but also to the virtual environment. The position and
orientation of every brick are available to the designer with high precision,
which makes event staging as simple as dragging-and-dropping modules into a
map. However, researchers do not have easy access to objects outside the
vehicle in on-road simulators. As a result, no on-road driving simulation
system to date considers the surrounding context outside the vehicle for event
staging, which limits the range of applicable studies. To replicate in-lab
simulation on on-road platforms, access to the out-of-vehicle surrounding
context is crucial. %We define the design process that considers running the same study on both in-lab and on-road simulators as \textit{cross-environment study design}.

In this paper, we describe a novel driving simulation infrastructure called
Portobello, which enables \textit{platform portability} in virtual driving
simulation by incorporating localization technology and software from robotics.
We define \textit{platform portability} as the ability to run the same study on
different (in-lab vs. on-road) platforms, an approach which we refer to as the
twinning of studies. For this demonstration, we extended XR-OOM, a state-of-art
XR driving simulation \cite{GoedickeXROOM}, to support on-road, location-based
event staging (see \autoref{fig:xroom_system}). To validate Portobello's
\textit{platform portability}, we developed a within-subjects
crosswalk-cooperation study (\textit{N}=32) to be run on both an in-lab
fixed-based vehicle chassis driving simulator and the on-road driving simulator
built on top of the Portobello system \cite{walch2020crosswalk}. As part of
this work, we investigated how the different simulation platforms may affect
the design process and results of user studies. The primary contribution of
this work is the technical infrastructure system of Portobello, as validated by
the proof-of-concept study. In addition, we provide a definition of
platform-portability in virtual reality driving simulation and contribute
insights into the process needed to develop twinned studies whose deployment is
intended across multiple platforms. Finally, we demonstrate the relative
strengths of different study platforms in the course of automotive research. %We subsequently reflect upon the study design process and survey results to share insights about the on-road simulator system design, cross-environment study design, and platform-driven influences on study results. 

\section{Related Work}

Driving simulation platforms are intended to be proxy environments that enable
researchers to conduct studies where real-world experiments are dangerous or
impossible. The standard for such platforms is face validity
\cite{dole2019face}: when participants take a simulation seriously, researchers
can have greater faith that the study's results will be applicable to the real
world. It is more important that the simulation allows participants to behave
as if they are in a realistic setting than it is for the simulation itself to
replicate reality in fine detail. In-lab and on-road simulators provide
different approximations of driving scenarios.

\subsection{In-Lab Driving Simulators}
In-lab driving simulators are used to test interactions between drivers and the
vehicle, the driving environment, and other in-world
agents~\cite{10.1145/3534617}; simulation allows researchers to observe
in-vehicle behaviors safely, without physical risk to study participants
\cite{tateyama_observation_2010}. Simulations can be implemented with low
fidelity but still provide insights into how drivers will react sitting in a
real vehicle \cite{colley2021swivr}. When used to simulate outdoor
environments, in-lab driving simulators can be used to test the usage of augmented reality (AR) on roads~\cite{tonnis2005experimental}. In-lab simulators can also be used to test how other road users, such as pedestrians and cyclists, interact with autonomous vehicles \cite{hou2020autonomous, mahadevan2019av}. However, simulator sickness remains a large risk for in-lab driving simulators because a
user's vestibular senses do not align with their visual senses when taking part
in a simulation \cite{bertolini2016moving}. Researchers have tried to address
this with methods such as aligning vehicular motion with VR content
\cite{cho_roadvr_2020} or simulating movement~\cite{10.1145/3494968,
    10.1145/3543174.3545252}, but even the highest-end simulators replicate a
fraction of the forces felt in normal on-road driving \cite{greenberg2003effect}.

\subsection{XR On-Road Driving Simulators}
\label{friends_of_xroom}
XR technology enables simulating AV driving either by allowing users to move through completely virtual spaces with real vehicle dynamics or by overlaying virtual objects on top of real-time video footage of the surroundings to increase the immersion of pre-programmed interfaces or interactions \cite{BringingSimulationtoLife, yeo2019maxim, GoedickeXROOM, riegler2020research, blissing2013augmented, HockCarVR, goedicke2018vr, sawano2005car}. Recently, XR systems have been deployed on-road to take advantage of the realistic road environment and vehicle dynamics. The XR-OOM system developed by \citet{GoedickeXROOM} employed an XR headset for drivers to drive through virtual, external obstacles in a parking lot. The MAXIM system developed by \citet{yeo2019maxim} utilized a virtual reality headset coupled with 360$^{\circ}$  cameras for subjects to experience an autonomous virtual vehicle situated in a live environment created from live streamed 360$^{\circ}$ videos. %\citet{ghiurau_arcar_2020} created a prototype of an XR experience where a driver-controlled a moving vehicle and tested it on roads closed to traffic. 
\citet{ARCAR2020} showed a proof-of-concept headset-based XR driving experience revealing that original equipment manufacturers (OEMs) such as Volvo use such technology.
Finally, \citet{PassengXR2022} presented PassengXR, an open-source toolkit to create passenger XR experiences. While providing XR experiences, they did not compare their system to an in-lab simulator.

Although all the systems mentioned above track the vehicles' dynamics to
accurately position virtual objects related to the participant, they do not
natively support high-precision interaction staging based on the surroundings
outside of the vehicle, which we refer to as \textit{surrounding context} in
this paper. In essence, previous approaches are mostly concerned with aligning
virtual and physical motions, such as CarVR, but not the worlds themselves
\cite{HockCarVR}. The authors of PassengXR introduced a hypothetical
application that requires high-precision alignment between virtual and physical worlds, where passengers on a bus tour could view AR-style information
overlaid on historic buildings outside the vehicle, but their demonstration was
still carried out indoors without discussing how feasible it was to implement
such application on-road \cite{PassengXR2022}. Potential issues for designing
such applications with existing platforms are two-fold. First, current systems
make it challenging for designers to stage interactions, where they need to
manually locate event trigger positions (in the bus tour example, buildings' coordinates
either in GPS coordinates or relative coordinates in the vehicle frame) and
program corresponding AR information boards to be at those precise coordinates.
Second, the GPS-based tracking system may not provide enough accuracy for
small-scale interaction, especially in cities where buildings can disrupt GPS signals. The authors for PassengXR commented that their "approach prioritizes perception of motion over location accuracy," which is not best suited for location-based XR experience \cite{PassengXR2022}. These hardships limit the capability of these on-road
systems to act as participant testing platforms compared to traditional in-lab
simulators. Most autonomous driving studies require sufficient staging and
surrounding context \cite{ColleyFeedbackStrategies}. For example,
driver-to-driver communications mostly happen at intersections, and
pedestrian-vehicle interactions usually occur at crosswalks
\cite{houtenbos2009interacting, lemmer2020driver, schneemann2016analyzing,
    krome2019people}. Our work makes high prevision interaction staging possible
while keeping the design process intuitive.

\subsection{Platform Portability}
Previous research has focused on how to replicate on-road scenarios in in-lab simulators, which is crucial when studying problems that are dangerous to experiment on the road, such as near-collision scenarios and passive rail level crossing \cite{demmel2011using, larue2018validation}. However, compromises are necessary to compensate for the lack of motion and sensory cues in in-lab simulators, and little research has been conducted to reduce the performance gap between on-the-road and in-simulators \cite{green2005driving}. As such, merely pursuing statistical significance with in-lab simulators may result in overlooking issues of practical relevance in real-world contexts. \cite{green2005driving}.

Because in-lab and on-road driving simulation environments offer different strengths when it comes to control and realism of driving scenarios, it can be desirable to run the same study in both when possible--- an approach we call twinning of studies --- to understand how study results from different environments relate to each other. \citet{hammel2002verbal} found that, when they replicated an on-road study in a fixed-based simulator, participants' eye-scanning behavioral patterns were similar, which demonstrates fix-based simulators' ecological validity.
In a systematic review of validation studies featuring comparisons of driving simulation and on-road driving between 1977 and 2017, authors \citet{wynne2019systematic} found only 44 validation studies comparing simulation to real driving. 
This is out of the 21,312 found by the same researchers to be English-language publications of original research having to do with driving simulation. Such studies are so rare that the 44 represent less than 0.25\% of the published driving simulation research surveyed by \citet{wynne2019systematic}. They report that "There was little consistency in the dependent measures used to assess differences between the simulator and on-road drive...Of particular concern is the fact that only half of the driving simulators were found to be valid and some were valid for one measure but not others." They note that since policy, legislation, and training are built off of simulator studies, a better understanding of which aspects of simulated studies are likely to carry over to real road conditions, and which are not critical \cite{wynne2019systematic}.

Frequently, we believe, the lack of validation studies is due to the significant challenge of creating "twinned" studies in both environments. The advent of on-road mixed reality simulation \cite{baltodano2015rrads, yeo2019maxim, yeo2020toward, GoedickeXROOM, PassengXR2022} makes it possible to bridge the divide using software events in the real world. However, no system has yet ported the same study course, code, and event design from one environment to the other. By making it possible to port studies developed for in-lab simulators to be run on-road--what we call \textit{platform 
portability}--we improve the ability for automotive researchers to extend their
in-lab studies to the real road, and thereby improve the validity of simulation
research. This would improve the ability of researchers to validate their
simulation studies in on-road environments, as recommended by
\citet{wynne2019systematic}: "Ideally this would see authors report empirical
validation evidence for their own simulator, and not relying on other
simulators as support for validity. Even if modeled on a previously validated
simulator, each set-up is unique and should be validated for those
specifications."

% THERE IS A GAP. Not many studies translate in-lab studies to on-road (because not so many ways to do on road). Some studies compare platforms, e.g. MAXIM to validate platforms. No studies porting the same course code from one environment to the other--this is the first.

%\subsection{Wizard-of-Oz in Driving Simulators}
% Wizard-of-Oz is a safe and efficient deployment method to test autonomous driving applications. In Wizard-of-Oz studies, participants interact with automated systems that are secretly operated by researchers, the wizards. In driving simulators, researchers can use Wizard-of-Oz to easily create an illusion of an autonomous driving system \cite{baltodano2015rrads}.  

% \subsection{Comparison between in-lab and On-road Simulators}
% - all system comparison papers focus on system comparison (motion sickness, presence)

\section{Systems}
\begin{figure}[ht]
  \centering
  \includegraphics[width=\columnwidth]{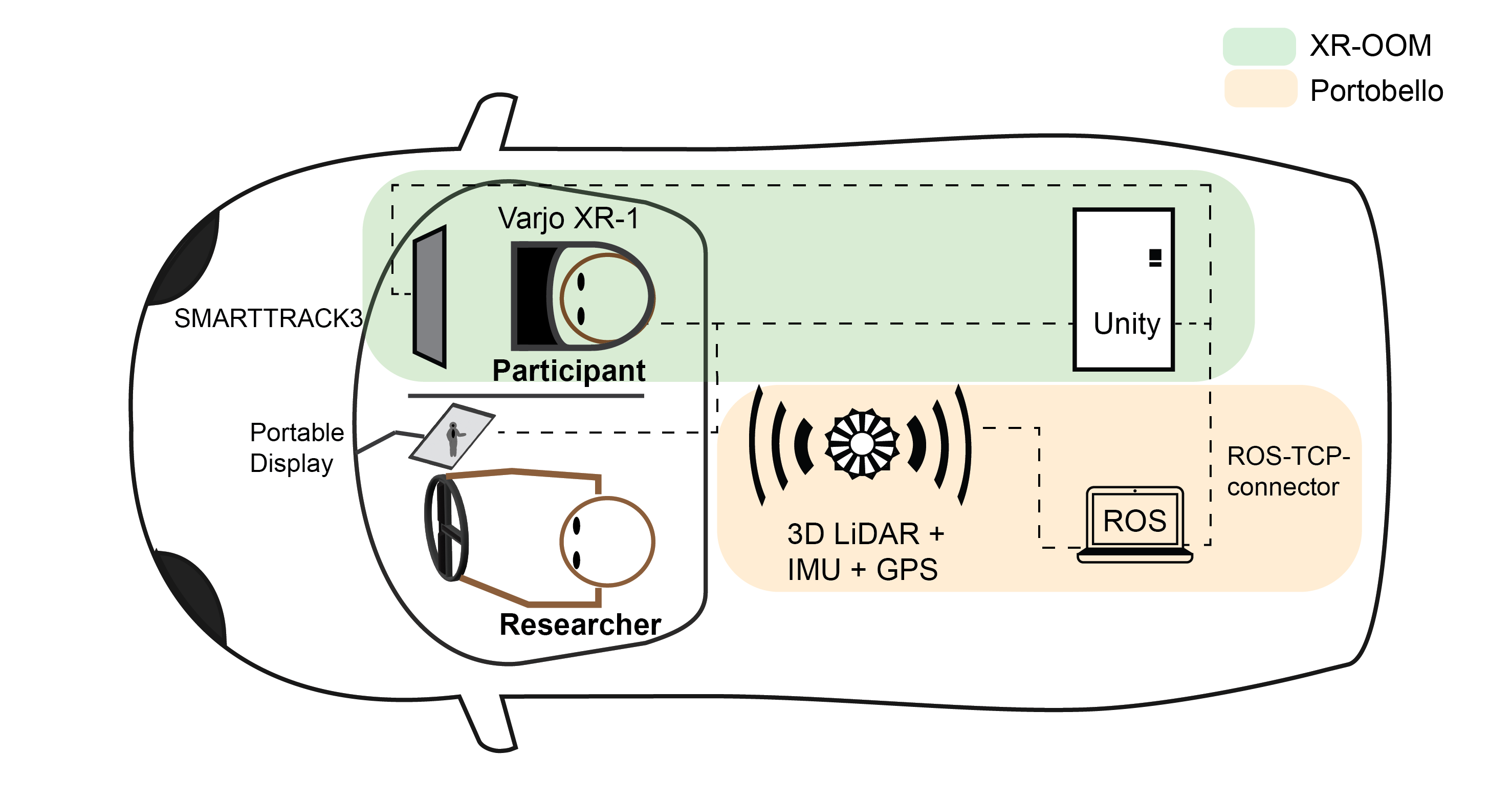}
  \caption{The Portobello system uses a LiDAR-based navigation stack to localize the vehicle's runtime position within a given map. The location information is transmitted to the Unity Desktop through ROS. The black dotted line indicates a virtual divider appearing only in the headset.}
  \label{fig:extended-xroom}
  \Description{This figure shows the schematic overview of Portobello. A top-down view of the vehicle is given. The schematic view shows the researcher sitting in the driver's seat, and the participant wearing the Varjo XR 1 headset in the passenger's seat. A mini-display is mounted next to the driver displaying the headset's view. The SMARTTRACK3 system is attached to where usually the glove compartment is located. The LiDAR connection is visualized and it is shown that this is connected to Unity via ROS. The equipment that is part of the XR-OOM system is highlighted in green, and the equipment that is part of the Portobello infrastructure is highlighted in orange.}
\end{figure}
In this work, we present a study we developed meant to run on two driving simulator environments: a lab-based driving simulator and an on-road XR driving simulator equipped with our Portobello system. Here, we recap the features of both environments and introduce the key adjustments made to accommodate the Portobello system.

\subsection{In-Lab Driving Simulator}
Our in-lab fixed-base driving simulator features a modified Fiat 500 in front
of three projector screens (see \autoref{fig:teaser}). The projector screens
cover participants' visual field when they sit in either the driver's seat or
the passenger's. The three projectors are DLP-based and can produce an image
with low latency on the projector screens. The projectors are connected to the
computer over HDMI and use the TripleHead2Go to split one DisplayPort signal
into the three outputs.

The vehicle is coated with non-reflective material to reduce the backscatter
onto the projector screens, increasing the contrast of the projector screens.
ButtKicker haptic transducers are installed under the front seats to provide
realistic tactile feedback from road noise and the engine. The simulation
software is run by an Alienware Area-51 R4 computer with two NVIDIA GTX1080 in
SLI. The vehicle's side mirrors are small digital displays rendered by the same
computer.

In our simulation software, the simulated vehicle is extended out of the Genivi Vehicle Simulator, which has been widely used in driving simulation studies \cite{genivi}. The simulated vehicle uses standard Unity physics wheel colliders with a built-in engine simulation. To enable autonomous driving, we replace the original steering wheel input with a waypoint-based navigation system.

\subsection{On-Road XR System}
\label{sec::onroad system}

To enable cross-environment study deployment, we use Portobello with the
XR-OOM system designed by \citet{GoedickeXROOM}. In the original XR-OOM system,
tracking and positioning of virtual objects are managed by a ZED 2 camera (for
visual-SLAM) and the ART
SMARTTRACK3\footnote{\url{https://ar-tracking.com/products/tracking-systems/smarttrack3/},
  accessed Jan 20, 2023} (for headset tracking within the vehicle)
\cite{GoedickeXROOM}. An onboard desktop running the Unity 3D game engine in
version 2020.3.26f1 overlays virtual objects on top of the "passed-through"
video of surroundings in the XR headset.

\subsection{Portobello System}
In the Portobello system, we use a LiDAR-based navigation system on the car
roof driven by the Robot Operating System (ROS 1 Noetic) \cite{quigley2009ros};
this replaces the XR-OOM's ZED 2 camera in front of the vehicle. The Portobello
navigation system features an Ouster OS-1 3D LiDAR with a built-in Inertial
Measurement Unit (IMU) and a ZED-F9R u-Blox GPS module. The communication
between the navigation system and the Unity Desktop is managed through the
ROS-TCP-Connector\footnote{\url{https://github.com/Unity-Technologies/ROS-TCP-Connector},
  accessed Oct 20, 2022} provided by Unity \cite{juliani2018unity}.

\textit{Platform portability} also drives augmentations to the XR-OOM software structure. Whereas XR-OOM uses real-time visual SLAM to compute short-term vehicle odometry, the Portobello LiDAR system enables global vehicle localization within a given map. As a result, virtual objects' positions are no longer associated with the vehicle position directly as its children. Rather, the vehicle and virtual objects share a common parent --- the world frame --- which is introduced by a map of the environment. Instead of staging virtual events around the vehicle, as is done in XR-OOM (and in all the other XR-based driving simulation systems mentioned in \cite{ARCAR2020}), Portobello can stage virtual events in a static shared map through which the vehicle drives. As Portobello replaces the car-centered reference frame with a map-based global frame, out-of-vehicle virtual objects remain fixed with respect to the map instead of to the arbitrary starting position of the vehicle. We detail the map generation process and staging process in Section~\ref{sec::portability}.

At runtime, Portobello's LiDAR-based navigation system updates the position of
a virtual vehicle in Unity. The relative position and orientation between the
virtual vehicle and the participant's headset are managed by SMARTTRACK3. The
virtual vehicle has the same shape as our research vehicle and is aligned with
the research vehicle throughout the ride. To simulate proper depth ordering,
the virtual vehicle is transparent with the alpha clipping option enabled. By
acting as a cutout shader, the virtual vehicle occludes virtual objects
outside. From the participants' view, the virtual objects are occluded by the
real car they sit in. As we are rendering virtual events in XR over a large
area (half of an island), we cap the headset's maximum rendering distance to
45 meters in order to improve motion parallax.

From a system design perspective, the computer running Portobello with ROS is a
separate computer from the computer running XR-OOM with Unity. In essence, the
communication between the ROS localization algorithm and Unity is achieved
through the ROS transform package (TF), which is a hierarchical tree structure
that tracks the relative position of multiple coordinate frames (map, LiDAR,
etc.). These coordinate frames can be accessed from Unity as game objects.
Isolating the system on a hardware level allows researchers to develop ROS and
Unity in their own environments and for one driving simulator to be swapped out
for another. 
 Designers can focus on designing studies by placing objects in the course map in Unity, rather than being concerned with low-level ROS localization of objects.
Another benefit of this practice is that the Portobello system consumes no computational resources in the original on-road platform at run time. 
The LiDAR-generated point-cloud map is rendered into Unity in the design phase. At runtime, the on-road platform computer imports the Unity map at start-up, and does not require the additional computational resources that would be needed to manage the point cloud map data. (This does mean that the Unity map might be missing physical features that change between the design phase and runtime.)

\section{Enabling Portability of Study Design Using Common Models}
\label{sec::portability}
To run twinned studies across platforms, we must keep portability in mind during the design of the study. We outline the necessary components to guarantee equivalent performance in cross-environment twinned studies and detail our system pipeline to showcase the components' connections using our system.

\subsection{Course Design}
On-road simulators are limited by real-world road infrastructures. Based on the
study focus, researchers should carefully consider test routes to ensure
efficiency and reproducibility.

As our study focused on interactions at crosswalks, we chose the southern loop
of Roosevelt Island as our study course. Most of the 0.9-mile long route
is single-track and has a high density of crosswalks, 15 in total. The drive
takes about 8 minutes. The route has no traffic lights and overlaps with two
bus lines.

\subsection{Map}
A map (model of the study area) is the starting point of event design, and its
precision and resolution greatly affect design complexity and quality. In
cross-environment twinned studies, a map is a bridge between simulation and the
real world, and it is also the shared common ground on every simulator
platform.

To generate a high-quality map, we used the LiDAR-based navigation system to
scan the entire test area. Specifically, we ran the real-time LiDAR-inertial
odometry package (LIO-SAM) developed by \citet{liosam2020shan} to create a
true-to-scale point cloud mapping of the study area. We drove through the
testing route multiple times to ensure loop closure. The resulting map is a
monochromatic digital twin that includes over one million points and captures
the test area in fine detail. Researchers can manually contextualize the point
cloud map in Unity with colored assets and use the map as the background in
in-lab simulators. %We skipped this step since we already have a digital twin of our study area from a previous study.
% Due to its size, it is difficult to import the point cloud as meshes to the simulation editor (Unity) directly. Hence, we used the Pcx package \footnote{https://github.com/keijiro/Pcx} to display the point cloud map with customized shaders. 
% By default, both Unity and ROS use the metric system. However, Unity employs the left-hand coordinate system, whereas ROS employs the right-hand coordinate system, which makes affine transformations necessary to match the point cloud map with the real world.

\subsection{Event Design}
Staging events along the planned course requires researchers to consider two
major questions: where and when events occur. Staging is relatively easy for
in-lab simulations, where agents' movement and speed profiles can be carefully
controlled to guarantee timing and location. In this section, we discuss how we
stage events using on-road simulators.

\subsubsection{Planned Events}

\paragraph{Where?} A one-to-one scaled map is necessary to plan events for on-road simulations
because the vehicle will drive through the real world during the study. Any
scaling or shifting on the map will cause significant errors at runtime. With a
loaded digital twin in Unity, researchers can drag and drop virtual agents and
objects just like they would for in-lab simulators.
\paragraph{When?} Timing of events for on-road simulators can be controlled through the placement
of collision-based triggers. Triggers in Unity are colliders that trigger
events upon external contact. For example, invisible triggers can be placed at
some distance $x$ in front of a virtual traffic light. Once the vehicle
collides with the trigger, the traffic light starts changing colors, and
participants should react accordingly. The distance $x$ governs the start of
the interaction, which essentially affects the maximum response time for
participants.

Of course, a vehicle on-road cannot "collide" with the virtual collider in
simulation. In our system, the LiDAR-based navigation system synchronizes the
position of the real vehicle with a virtual vehicle in the digital twin through
the $hdl\_localization$ and \textit{ROS-TCP-Connector} packages in real-time
\cite{koide2019portable}. 
(By employing the $hdl_localization$ algorithm, the offset between the vehicle's actual and estimated locations is maintained within a 0.2-meter range.)
As the real vehicle drives through the world, the
virtual vehicle simultaneously moves through the digital twin to trigger
planned events. Note that this is the same virtual vehicle mentioned in section
\ref{sec::onroad system} for proper depth ordering.

\begin{figure}[t]
  \centering
  \includegraphics[width=\columnwidth]{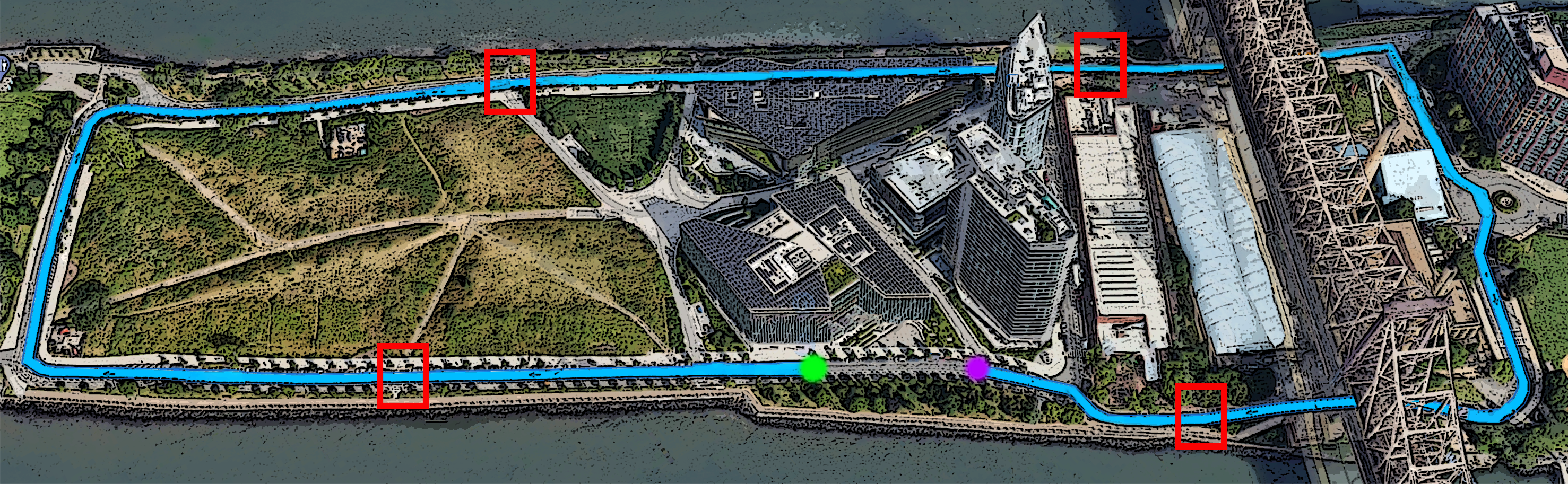}
  \caption{Our study area for on-road simulator on \anon{Roosevelt Island}. The pre-determined test route is highlighted in blue. The crosswalks with staged interactions are highlighted in the red bounding boxes. The start and end locations are denoted by the green and purple dots, respectively.}
  \label{fig:rooseveltIsland}
  \Description{This figure shows a top-down satellite image of Anonymous. The pre-determined test route is highlighted in blue. It spans an approximate rectangle.}
\end{figure}

\subsubsection{Unplanned Events}
We define events outside the simulation, which researchers have no control
over, as \textit{unplanned events}. In in-lab simulations, unplanned events are
rare and typically caused by system failures or external interruptions.
However, in on-road simulators, unplanned events are common and can even be
valuable for ecological validity; they let researchers know if their findings
are robust to real-world variation. Findings that are only true in the tightly
controlled environment of a study have little practical application.

Nevertheless, researchers must factor in potential unplanned events during the
design phase to ensure safety and preserve meaningful study results. Unplanned
events come in different forms, from unexpected appearances of pedestrians to
weather changes. For instance, in our study, we encountered the following
unplanned events: real pedestrians and geese crossing the street, other
vehicles passing the research vehicle from the bicycle lanes, and rain.

\subsection{Platform Measures}
Another aspect of \textit{platform portability} is whether researchers can
obtain the same set of measures from twinned studies.
%different platforms. 
While measurements should be equally attainable across all simulation
platforms, the characteristics of each simulator naturally encourage and
discourage different sets of measures.

\subsubsection{Behavioral Response}
Behavioral responses refer to the participants' elicited behavior during the
study. Examining behavioral responses is crucial when studying interactions between drivers, vehicles, and infrastructure \cite{ruscio2017collection}. For example, \citet{Jansen2023} was interested in differences in participants'
responses to different stimuli in automotive user interfaces. With
appropriate sensors, collecting behavioral responses in in-lab and on-road
simulators is possible.

\subsubsection{Performance Response}
\citet{paas1993efficiency} define performance as efficiency in completing tasks. We distinguish performance from behavioral responses based on the availability of ground truth. Researchers can collect performance responses during the study when participants are assigned tasks with general guidelines and standards. One example of performance response is the lateral vehicle position when the driver is distracted~\cite{kountouriotis2016leading}. While extra sensors might be needed for on-road simulators to obtain vehicle-related performance measures (e.g., vehicle speed, acceleration, or trajectories are not easily attainable in on-road simulators as they are in in-lab simulators), we do not anticipate significant challenges in obtaining performance responses in both in-lab and on-road simulators.

\subsubsection{Survey Response}
Surveys can be conducted through different devices (pen and paper, tablets) in
various formats (interviews, multiple choice, open-ended questions). In
portable study design, it is important to consider the timing of the survey. One
natural advantage of in-lab simulators over on-road simulators is the ability
to pause at any point of the study and prompt participants with questions in
situ \cite{de2012advantages}. On-road simulators cannot be paused easily, so surveys need to be planned so that participants can take them when it is safe to do so.

\subsection{Additional Instrumentation}
Detailed runtime recording of the environment is crucial for post-facto data
analysis, particularly of unplanned events, for both on-road and in-lab
simulators. Some measures need to be recorded differently in the different
platforms and translated. For example, geo-location data from the on-road
vehicle GPS must be correlated with the virtual world coordinates in the lab
simulator. Head orientation and gaze direction obtained from the XR headset in
the on-road simulator can be correlated with camera-tracked head-pose in the
in-lab driving simulator.

\section{Twinning of Studies}
As proof of concept that we can run the same study design in the lab and
on-road (twinning of studies), we conducted a within-subjects experiment with
\textit{N}=32 participants (under IRB\#\anon{1806008105}). We described the
cross-platform deployment of twinned studies and compared the differences in between. 
We counterbalanced the experiment conditions, where half of the
participants experienced the study in the indoor simulator first and in the
on-road simulator second, and the other 16 participants experienced the
simulators in the reverse order.
%https://www.psychologie.hhu.de/arbeitsgruppen/allgemeine-psychologie-und-arbeitspsychologie/gpower

We employed the in-lab and on-road driving simulators to run twinned Crosswalk
Cooperation studies, which we adapted from \citet{walch2020crosswalk}. In this
previous study, Walch et al. used an in-lab driving simulator to evaluate the
usability of a novel car UI and staged interactions around a driving loop. 
In this current work, we are not seeking to validate the results of the previously published study; we are not expecting or arguing that our study results would be the same. Rather, we are merely using this study design to evaluate the capacities and key influences of both systems.

\subsection{Study Setup}

\subsubsection{Protocol}
The experiment is a within-participants experiment design; each participant
experiences two study sessions in counterbalanced order. In one session,
participants experience the crosswalk cooperation study in the in-lab driving
simulator. Afterwards, they fill out a post-session questionnaire, which collects information on their experience with the simulator. In the other session, the participants are escorted to the curbside and experience the crosswalk cooperation study in the on-road driving simulator. Afterwards, they fill out the same post-session questionnaire. Finally, they fill out a post-study questionnaire, which collects information on their perceived differences between the two simulators.

During each simulation session, participants are seated in the front passenger seat and informed that they will experience automated driving: The vehicle will stop at all crosswalks automatically. The vehicle will proceed autonomously when road conditions are clear (e.g. crosswalks without virtual pedestrians) and will ask for input from the participant via a smartphone interface, on how to proceed in unclear situations (e.g. virtual pedestrians walking towards crosswalks). Specifically, the vehicle will ask, "Is now safe to proceed?" on the phone interface while waiting at the crosswalk; when the participant feels it is safe to proceed, they press the "Proceed" button, and the car resumes its predefined route. If the participants decide it is not safe to proceed, they should wait until it is safe to press the button. The researcher driving the vehicle monitors when the participant presses the button and manually proceeds with the course if it is safe.

\subsubsection{Differences in Simulator Setup}
As much as possible, we maintained identical setups for the twinned studies.
% the in-lab and on-road simulators. 
The key difference was that during the on-road simulation session, the
researcher driving the car was mindful of the actual road conditions before
proceeding with driving.

To maintain the autonomous driving narrative, the
researcher driving during the on-road simulation was masked by a black divider
in the video pass-through headset (shown in \autoref{fig:extended-xroom}). A
similar black divider was also installed in the in-lab simulator to maintain
setup parity. During pilot sessions, we found it difficult to disguise the
on-road vehicle as an AV due to differences in sound profile. In complex
on-road conditions, the sound of pressing the pedal and rotating the steering wheel
broke the illusion quickly. Thus, we decided to inform the participants of the
divider's purpose and that there was a real researcher in the car with them
operating the vehicle.

While a virtual map environment is required in in-lab simulation, the point
cloud map is not required for on-road simulation due to the benefit of the
video-pass-through headset. Therefore, after the event staging phase, we
disabled the point cloud rendering in Unity to save computation power.

% \subsubsection{Video Recording}

\subsubsection{Scenarios}
We recreated four scenarios from the original Crosswalk Cooperation study by
\citet{walch2020crosswalk}. Each scenario was engineered so that pedestrian
interactions would only happen at crosswalks. In each scenario, virtual
pedestrians interacted with each other on the sidewalks near the crosswalks. In
half of the scenarios, one pedestrian walked to the stop sign and crossed the
street after giving clear body language signals that they intended to cross
(looking left and right). In the other half of the scenarios, the pedestrian
stopped at or walked past the crosswalk. In our version of the Crosswalk
Cooperation study, we constrained our study area to the southern loop of
\anon{Roosevelt Island}.

\subsection{Participants}
Out of our 32 participants ranging from 20 to 47 years old (age: \textit{M} =
27.38 $\pm$ 5.92), 19 participants identified themselves as male, 11 as female,
and two as non-binary. Six participants had experience with AV simulations
and/or AV research, and five had experience with commercialized AVs (Tesla,
demos at car shows). The others had little experience with AVs.

\subsection{Study Measures}
After each session, participants filled out a post-session questionnaire, which
collected information on their experience with the simulator. After completing
both sessions and corresponding questionnaires, participants filled out a
the post-study questionnaire, which collected information on their perceived
differences between the two simulators.

We also recorded video and audio for all sessions run in both in-lab and
on-road driving simulations to investigate participants' behavioral responses. 
For the in-lab driving simulator, a go-pro camera was pointed towards the participant to
record their upper body. The simulated virtual environment was recorded using screen
recording software. For the on-road driving simulator, similar to the setup in
the in-lab simulator, a camera was mounted in the glove compartment to record
the participant's upper body. An additional camera was mounted near the rear
mirror facing forward to record the road condition ahead. The participant's
XR view (video pass-through with overlay) was also recorded using the Varjo Base software.
\section{Results and Analysis}
We analyzed the video footage and the questionnaires from both sessions with
the goal of investigating the influences of both driving simulators on the user
study experience and understanding what results from one platform predict for
results on the other. Our evaluation of Portobello is based on being able to run and gather comparable results from the studies on both the in-lab and on-road platforms and not on the originality, validity, or significance of the study itself.

\begin{figure}[ht]
    \centering
    \includegraphics[width=\columnwidth]{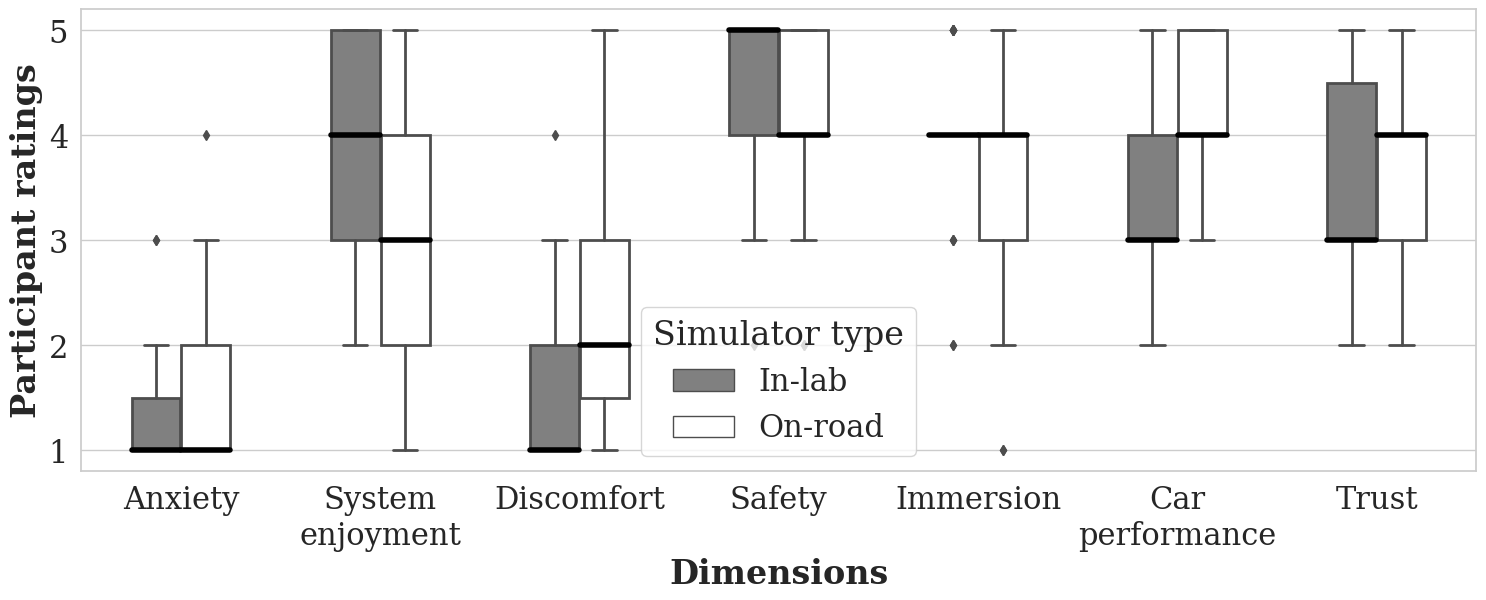}
    \caption{Participants' ratings of each simulator on a five-point Likert scale across seven distinct dimensions where 1=low and 5=high.}
    \label{fig:postdrive_stats}
    \Description{This figure shows the ratings of each simulator on a five-point scale across anxiety, system enjoyment, discomfort, safety, immersion, car performance, and trust. The actual values are given in the text.}
\end{figure}

\subsection{Study Results}
\subsubsection{Measures}
To compare the overall participant measures in the study on both simulators, we
asked the participants (\textit{N} = 32) to rate their feelings of anxiety,
safety, and trust on a 5-point Likert scale for each of the platforms. We
designed the questions based on the questionnaire used in
\citet{walch2020crosswalk}'s original Crosswalk Cooperation study. We ran a
Bayesian factor analysis on the captured measures with the null hypothesis that
there is no difference between the platforms \cite{kass1995bayes} with R
version 4.3.2 and the BayesFactor
package~\cite{moreyRicharddmoreyBayesFactor2022} using Jeffreys-Zellner-Siow
(JZS) priors, i.e., the default non-informative Jeffreys prior. Interpretations
were made according to \citet{jeffreys1998theory}. All packages were up to date
in November 2023. Since we have limited data, and we are not interested in how
different variables (ratings) interact with each other, we chose to report
single variate analysis over multivariate analysis. Nonetheless, our claims
hold under multivariate analysis as well. We will provide R script for both
analyses.
% For descriptive results, see \autoref{tab:_descr_stat}. 

One participant left multiple answers empty, so we dropped their results in the following analysis.

\paragraph{Anxiety}
More participants reported reduced anxiety with the in-lab simulator
(\textit{M} = 1.32 $\pm$ 0.60) than with the outdoor simulator (\textit{M} =
1.45 $\pm$ 0.72). We found \textit{moderate} evidence (\textit{BF} = 0.32) in
favor of the null model, suggesting that there is no significant difference in
the anxiety generated by the simulators.

\paragraph{Safety}
Participants considered the in-lab simulator (\textit{M} = 4.58 $\pm$ 0.76) safer than the on-road simulator (\textit{M} = 4.16 $\pm$ 0.86). We found
\textit{moderate} evidence (\textit{BF} = 3.53) against the null model,
suggesting a moderate difference in favor of the in-lab simulator.

This may have been because the in-lab simulator's roads did not involve any
real vehicles or people, e.g., P6 explained that "... there were more actual
obstacles...to take into account [in the outdoor simulator] whereas the in-lab
[simulator] had a preset number."

\paragraph{Trust}
Overall, participants reported their trust in the simulated autonomous driving to be higher in the on-road simulator (\textit{M} = 3.71 $\pm$ 0.86) than in the
in-lab simulator (\textit{M} = 3.52 $\pm$ 1.06). We found \textit{moderate}
evidence (\textit{BF} = 0.31) in favor of the null model, suggesting there is
no significant difference in trust.

The preference for the on-road simulator may have been influenced by people's
perception of each vehicle's performance. At the same time, this may have been
a breakdown in face validity. For example, P14 said, "One of the survey
questions asked 'how much do you trust this car' and I think I forgot to
pretend that this was an AI driving the car when answering that..."
Participants were informed during the on-road simulator session that a
researcher would be driving the car; their trust rating might have been an
indicator that they trusted the driver to obey local traffic laws, rather than
an indicator of their trust in the simulated autonomous driving, as we had
intended.

\subsubsection{Cooperation Behavior}
From recorded videos, we analyzed 32 participants' cooperative behaviors with
the vehicle at crosswalks. Video recordings from the same session were synchronized before the analysis. One researcher watched recordings for each participant and labeled their behaviors in terms of the timing of cooperation behaviors for each scenario. The researcher then noted down the behavioral changes, if any, for each participant between different simulators.

15 participants cooperated with the vehicle perfectly in both simulation platforms in all scenarios, where they waited until the pedestrians fully crossed the street or waited for clear non-crossing signals before instructing the vehicle to proceed. During the first crossing scenario, 11 participants instructed the vehicle to proceed while the pedestrian was about to cross; this led to six virtual collisions. One of these participants made the vehicle run over the same virtual pedestrian in both in-lab and on-road simulators. Three participants did not wait for any virtual pedestrian to cross in both in-lab and on-road simulators. Six participants who had cooperated perfectly in the first session made different decisions in the second session; they chose not to wait for the crossing participants when they believed it was safe, and one of them ended up running over the virtual pedestrian. One participant who ignored both crossing pedestrians in in-lab simulators waited for one of the crossing pedestrians in the on-road simulator. Overall, 14 participants made different decisions in the second session from the first session.

We feel compelled to point out that the fact that participants made poor
crossing decisions is not a sign that the study or the driving simulation
platforms were designed poorly; instead, it is precisely these sorts of outcomes that indicate the necessity for simulation platforms that enable studies with virtual pedestrians to be conducted prior to putting real pedestrians in harm's way.
% the point of these kinds of studies and platforms is the very reason why we need to run participant studies, and need simulation platforms to run them with. 
We do not expect this means that participants would run over real people in
subsequent tests with real cars, but this does point out that participants are aware that they are not exposed to real danger in driving simulators \cite{jamson2010validity}; impatience and lack of conscientiousness amongst some portion of the population are factors
that any cooperative autonomous driving system would have to account for.

\subsubsection{Ordering Effect}
While we counterbalanced our study, we noticed some differences in cooperation
behavior that may be attributed to the ordering of simulators. Of the six
participants who made mistakes in the first crossing scenario, four
participants were experiencing the on-road driving simulator. We hypothesize
that the on-road driving simulator is more overwhelming than the in-lab
simulator to familiarize the participants with the study setup. P28 mentioned
in their post-study questionnaire that "Visual noise in outdoor sim [made] task
completion more difficult but [was] more realistic in that regard." It is worth
mentioning that the three participants who made the mistake in the first
crossing scenario had limited (e.g., they had only their learner's permits) to
no driving experience.

\subsection{Simulation Evaluation Results}

\subsubsection{Simulator Measures}
To compare the overall experience between the two simulators, we also asked
participants to rate their feelings about car performance, system enjoyment,
discomfort, and immersion on a 5-point Likert scale. These questions were based
on questionnaires used during the validation process of \citet{GoedickeXROOM}'s
XR-OOM system. We again ran a Bayesian factor analysis on the captured measures
with the null hypothesis that there was no difference between the platforms
\cite{kass1995bayes}. % with R version 4.2.3 and the BayesFactor package~\cite{moreyRicharddmoreyBayesFactor2022} using JZS priors.

\paragraph{Car Performance}
Participants considered the autonomous vehicle in the on-road simulator
(\textit{M} = 4.35 $\pm$ 0.71) to perform better than in the in-lab simulator
(\textit{M} = 3.42 $\pm$ 0.96). We found \textit{extreme} evidence (\textit{BF
    = 1.67e+04}) against the null model, suggesting a significant difference in
favor of the on-road simulator. Participants felt that the in-lab car simulator
did not appear to drive smoothly and stopped rather abruptly at times and thus
thought that the driving felt more natural in the on-road simulator.

\paragraph{System Enjoyment}
More participants reported increased levels of system enjoyment with the in-lab
simulator (\textit{M} = 3.84 $\pm$ 0.97) than with the outdoor simulator
(\textit{M} = 3.03 $\pm$ 0.98). We found \textit{extreme} evidence (\textit{BF}
= 225.00) against the null model, suggesting a significant difference in favor
of the in-lab simulator. Participants seemed to have preferred the in-lab
simulator because of the graphics quality and comfort. For example, when asked
to describe their experience with both simulators in the post-study
questionnaire, P14 said, "The pedestrians seemed to "appear out of nowhere" in
the [on-road simulator], whereas it seemed like they were always part of the
scenery in the [in-lab] simulator (i.e., came into view naturally in the
simulator). ... [The] turns in the [in-lab simulator's] road felt
unnatural/like the scenery was clicked and dragged in front of my eyes, instead
of me moving through the scenery."

\paragraph{Discomfort}
Participants reported less discomfort with the in-lab simulator (\textit{M} =
1.71 $\pm$ 0.86) than the on-road simulator (\textit{M} = 2.35 $\pm$ 1.14). We
found \textit{anecdotal} evidence (\textit{BF = 2.24}) against the null model,
suggesting a mild difference in favor of the in-lab simulator. For the in-lab
simulator, discomfort mainly arose from the unrealistic vehicle dynamics. For
the on-road simulator, many comments were related specifically to the XR
headset (\textit{n} = 11), which people found heavy and uncomfortable. The
occasional misalignment of virtual objects caused by bumps in the road also
induced a considerable amount of motion discomfort. For example, P13 said,
"...both [simulators] cause some discomfort, but I think the outdoor one is
more uncomfortable due to the [pass-through] being very [shaky] and more
motion-sickness-inducing."

% P5: Using the VR headset, I found it difficult to see where the people were standing. This was especially true as the car bounced and the people in the scene got displaced.

% P6: The in-lab simulator caused more discomfort although both only caused slight discomfort. I think the in-lab caused more since I wasn’t actually moving compared to the on road. I think if I was in either system for extended periods of time, it would lead to more discomfort.

% P7: The in-lab simulation caused a bit more discomfort due to the car’s rapid acceleration/deacceleration for crosswalks and being unsure if the car was even going to stop. The discomfort in the outdoor simulation was more related to not knowing exactly where the AR objects were in the real world, especially when it came to their position in relation to the crosswalk. 

% P9: No significant difference in experience. Would say a bit more relaxed when doing experiment outdoor (no sure if it is because the in-lab experiment has been done previously, making myself more familiar with the following up outdoor one).

% P13: They both cause some discomfort, but I think the outdoor one is more uncomfortable due to the passthrough being very shakey and more motion-sickness-inducing. 

% P6: "- On-road there were more actual obstacles/people/surroundings to take into account whereas the in-lab had a preset number
% - in-lab car movement feeling was not as close to the outdoor real car movement"

\begin{figure*}[ht]
    \centering
    \includegraphics[width=\textwidth]{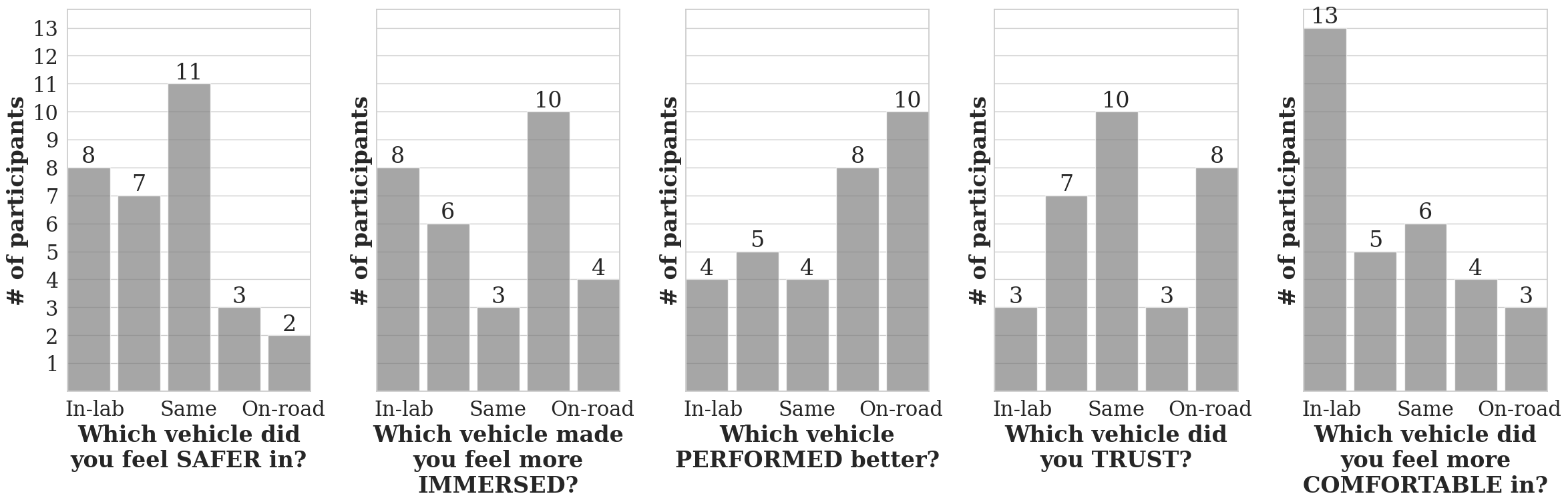}
    \caption{After experiencing both simulators, participants also directly compared the in-lab simulator and the on-road simulator on the same Likert scale. Participants' responses are shown as histograms.}
    \label{fig:post_study_hist}
    \Description{This figure shows the histograms of the participants' responses to different questions comparing the two simulators on 5-point Likert scales. The questions were: Which vehicle did you feel safer in? Here, the tendency was clear towards the in-lab simulator. Which vehicle made you feel more immersed?  Here, the tendency wasn't that clear with 8 claiming totally in-lab but 10 stating more on-road. Which vehicle performed better?  Here, the tendency was clear toward the on-road simulator. Which vehicle did you trust? Here, 10 participants stated ``same'' but also 8 stated on-road. Which vehicle did you feel more comfortable in?  Here, the tendency was clear towards the in-lab simulator.}
\end{figure*}

\paragraph{Immersion}
As shown in \autoref{fig:postdrive_stats}, participants considered the in-lab
simulator (\textit{M} = 4.00 $\pm$ 0.77) to be more immersive than the on-road
simulator (\textit{M} = 3.74 $\pm$ 1.06). We found \textit{anecdotal} evidence
(\textit{BF} = 0.0.40) in favor of the null model. However, in the post-study
questionnaire, when both simulators were presented on the same Likert scale,
the numbers of participants in favor of either simulator were the same.
\autoref{fig:post_study_hist} shows that 14 participants thought the in-lab
simulator was more immersive, and 14 thought the on-road simulator was more
immersive. Three participants considered both simulators equally immersive.

Two participants reported verbally during the on-road session that it was
difficult to distinguish virtual from real pedestrians.

Many participants reported difficulty with the weight and technical maturity of
the XR headset in the outdoor simulator, which impeded their attention and may
have contributed to its lower immersion rating. P8 said, "headset jitter made
visuals blurry, which impaired decision-making/attention." P2 said, "[Putting]
something on my head is so uncomfortable. I couldn't [focus on] the view or be
relaxed. The [in-lab] simulator is not so real[,] but I could be so relaxed."

\subsubsection{In simulator behavior}
We weighed the observed and reported behavior of participants to assess
differences in behavior introduced by each platform.

\paragraph{Natural Head Motions}
We noticed that some participants' head pose motion patterns were different
when experiencing the two simulation platforms. When sitting in the in-lab
static simulator, participants tilted their heads mostly during staged events
to track the motion of virtual pedestrians. For the rest of the ride, they
faced forward. However, for the on-road simulator, the participants' head
motions were more varied, and participants naturally looked to view the
surroundings more often.

%\paragraph {Perceived Weight of Decisions}

\paragraph{Decision-making}

For the crosswalk cooperation study, the key measure was the participant's
decision-making around whether the vehicle should go or not. Participants
reflected that the level of complexity and severity in decision-making was
greater in the on-road simulator. The staged interactions were identical in both
sessions, but researchers had less control over surrounding factors during the
on-road sessions. Other road users, including real pedestrians, other vehicles,
and geese native to \anon{Roosevelt Island} complicated the staged scenarios.

P5 said, "I think that in the on-road simulator, I was a little more nervous
because real people were on the street. The in-lab simulator does not deal with
real people, so any mistake I make does not have as much weight." P28 said,
"Outdoor provides [a] more generally immersive feeling and is the only one in
which I can realistically feel unsafe, which is a positive in terms of
validity. Visual noise in outdoor sim makes task completion more difficult but
is more realistic in that regard." P31 said, "I felt anxious after saying to
proceed and wondering how the pedestrians would move afterwards." In contrast,
P11 reflected: "Perhaps due to its nature of being [an in-lab] simulator, I
felt at ease at all times." P20 reflected: "The [polished nature] of the
[in-lab] simulator makes the experience [feel] more entertaining/performative
rather than realistic. ... The [on-road] simulator feels more like a functional
approximation of an actual autonomous driving car, while the indoor simulator
feels more like a fun, polished experience." P21 reflected: "I felt much
safer/less anxious in the indoor simulator which also probably means it was
less realistic." P29 reflected: "[The] indoor [simulator] seems to have lower
stakes, even though it was the same virtual people.

\section{Technical Validation}
Since the Portobello system is an infrastructure meant to be used in
conjunction with existing platforms, the technical performance largely depends
on the system on top of Portobello. Therefore, we investigated the change in
performance of the on-road platform after adapting Portobello. Rendering a
total of 12 virtual pedestrians, the headset ran at 60 FPS consistently with a
display latency of around 35ms, which was on par with the original XR-OOM
system. The localization frequency was 10 Hz, limited by the 10Hz LiDAR. The
fact that there is no change in performance is expected since the Portobello
system operates on a computer separated from the original XR-OOM system.

\section{Discussion}
\subsection{LiDAR-based vs. GPS/IMU-based systems}
Our LiDAR-based Portobello system resolves the two challenges posed by traditional GPS/IMU-based systems in surrounding context-based interaction staging: localization accuracy and design hardships. LiDAR-based localization systems work reliably in cities where buildings serve as landmarks instead of GPS signal blockers.  A point cloud 3D map of the environment generated from LiDAR-based SLAM algorithm saves designers from staging using hardcoded coordinates, and simplifies the design process to drag-and-drop within the map. 
Lastly, we want to point out that the sensors are not mutually exclusive. We can fuse in GPS data as an additional data source into the LiDAR-based algorithm if necessary.

\subsection{Platform Portability Challenges}
The primary goal of this research effort is to establish a proof-of-concept demonstration of \textit{platform portability} through our Portobello system. \textit{Platform portability} is necessary to run twinned studies across different platforms, which is desirable because the in-lab simulator can help establish causal differences across experimental conditions using well-controlled studies, and the on-road simulator can help validate the ecological validity of such study results when the same study conditions are moved into the less-controlled environment of the real world. By incorporating robotics mapping, sensing, and localization capability in the Portobello system, we can set up twinned studies to run in different environments. We believe that this is of relevance to any of the simulators discussed in Section \ref{friends_of_xroom}, which could be deployed on top of Portobello. 

\subsubsection{Randomness in the Wild}
For on-road simulators, randomness persists throughout the entire study. For example, during mapping, we generated a snapshot of the test area. While major landmarks such as buildings and land topology will not change significantly over time, the map also captures transient objects (e.g., parked cars). Such randomness may cause inaccuracy in real-time localization. 

During the study, unplanned events were the most salient form of randomness. Unplanned occurrences and interruptions from the real world may increase immersion. P19 reported that "the extra people and cars in the outdoor study made the experience feel more immersive and interesting." However, randomness also brings concerns regarding study reproducibility. We tried to eliminate the co-occurrence of planned and unplanned events by staging events in less populated areas. In general, we encourage researchers to plan for all possible unplanned events during the study design phase.

\subsubsection{Event Timing} \label{sec::timing}
One major challenge we faced with the on-road simulator is the trigger and timing of staged events, and we foresee such a problem persisting in future similar studies. For in-lab simulators, vehicle speed and travel distances can be coded in detail. For on-road simulators, it can be difficult to maintain the same speed curve as in-lab simulators due to obstacles and unplanned events. In our study, we expect the vehicle to stop at the crosswalk simultaneously as a virtual pedestrian reaches the stop sign on the sidewalk. The researcher who operates the car has access to a mini-display monitoring the location of the virtual pedestrians and adjusting vehicle speed accordingly. However, we have noticed that participants made different decisions across platforms due to event timing differences.

The timing misalignment between the simulators is the natural consequence of the intentional difference between running studies in a controlled environment (the lab) and an uncontrolled environment (the real world). We are arguing that it is desirable to run both kinds of studies and that it is easier to do this if \textit{platform portability} exists. The in-lab simulator is more suitable for quantitative analysis, and the on-road simulator is more suitable for qualitative analysis of the factors that complicate the outcomes learned from the more controlled in-lab simulator.

\subsection{Platform Effects}
Our study results were intended to help us understand \textit{whether and how} our twinned studies were the same across the two platforms and to help us understand the differences across the platforms, a comparison made possible by the Portobello system. For research purposes, it would be best if the study results between the two platforms were similar (i.e., that the platform effects were negligible), or at least that the results were biased consistently across the platforms (i.e., that the platform effects were predictable). 

From the study results, we can see that around half of the participants made similar decisions in both the in-lab simulator and the on-road system. Notably, one participant made the same mistake in each simulator. The fact that some participants made different decisions in different simulators indicates that participants' decision-making was not affected by their existing knowledge about the study. Even if they were aware of the pedestrians' behaviors from the first session, they took into account the event timing difference and made the most appropriate decision at the moment.

Some notable differences between the platforms were centered around the participant's decision-making behavior and their resulting trust in the autonomous driving system. In some ways, there were indications that the on-road simulator might have failed to maintain face validity for at least one of the participants; their trust rating might have shown their trust in the research driver rather than their trust in the simulated autonomous driving, as we had intended. On the other hand, many participants reported greater weight in the decision-making around whether the vehicle should go or not go in the on-road simulator, a sign that face validity is higher in the real-world environment.

While our participants favored the experience of the in-lab simulation platform on a whole, most of their complaints pertained to aspects of the XR system--the weight of the headset, the jitter in the display--which are likely to improve with advancing technology. It seems like the naturalism of the on-road environment is more likely to yield naturalistic behaviors than the in-vehicle environment, and hence this platform can help industrial and academic researchers better understand how people will engage with autonomous vehicle technologies in the real world.

\subsection{Limitations and Future Work}
Some limitations in the study results are inherent to a driving simulation study. In this section, we focus on limitations in the design and execution of studies using the XR driving simulation system augmented by Portobello that should be accounted for and discuss future developments that could improve such systems.

\subsubsection{Real-time adjustment of Depth Ordering}
Although we render the occlusion of virtual objects caused by the research vehicle (e.g. we do not render the pedestrians over the front pillar of the car), the current system does not provide the same occlusion for runtime dynamic objects. If a bus drives between the research vehicle and the location where the virtual pedestrians are supposed to appear, for example, participants would see the virtual pedestrians in front of the bus. To correct the depth order, future systems could use real-time LiDAR scans of the environment.

\subsubsection{Pedestrian Appearance}
We cap the maximum rendering distance for on-road virtual objects for technical reasons; distant virtual objects are less salient and require better alignment between the in-vehicle and out-of-vehicle reference frames to be placed believably in the mixed-reality view. The artifice of having pedestrians suddenly appear, however, may affect study results. One participant (P24) said, "I think being able to see the passengers from further away in the in-lab simulator made a big difference because, by the time I got to the intersection, it was easier to anticipate their movements." Future technology could improve the motion parallax issues, enabling longer rendering distances and smoother transitions when virtual objects approach the rendering threshold.

\subsubsection{Headset Discomfort}
Many participants complained about the bulkiness and narrow field-of-view of the headset. This platform-level discomfort was pronounced enough that it drowned out our ability to measure experiential aspects (system enjoyment, discomfort) of the autonomous driving scenario. While the weight and limitations of the XR headsets are beyond our control, we believe that anticipated advancements in XR headset technology are necessary to use these systems in experiments wherein the experiential aspects of automated driving are critical.  

\subsubsection{Simulating Autonomous Driving on Road}
In our current study, we informed the participants that there was an actual driver behind the scenes in the on-road simulator. The driver's maneuver sound easily breaks the AV illusion for participants who have previous experience with AV. Future research can benefit from disguising the driver by playing the recorded AV sound profile during acceleration and deceleration to cover the driver's maneuver sound. 

\subsubsection{Consistent Driver Performance}
While the same researcher operated the vehicle throughout the study, and attempted to maintain a consistent driving style from one run to the next, there were natural variations in the driving, in part in response to uncontrolled environmental factors. While it is not feasible nor desirable to force the driver to operate the vehicle the same way for all sessions on the road, future systems should collect data to enable later analysis of variance.

\section{Conclusion}
Driving simulations can be used to create scenarios for driving interactions, which enable researchers to better understand how people will behave and respond to future driving scenarios. In this work, we present the Portobello system, an on-road driving simulation infrastructure that enables \textit{platform portability}. By advancing the capabilities of driving simulators, we can better anticipate what aspects of driving interaction will work well or poorly. 

 This paper outlines the first-ever deployment of twinned studies across in-lab and on-road simulators. We found that participants preferred the experience of the in-lab simulator but displayed more natural head movements in the on-road simulator; they also reported that the decisions made in the on-road system carried more weight. Based on our findings, we suggest researchers working in driving simulations also take the twinning of studies approach: they should first run studies within a controlled, in-lab environment to collect statistical measures and form hypotheses and then port their studies to a less-controlled, on-road simulator and test their hypotheses in a more complex, realistic environment.

 This experiment looking at the platform-driven influences on study outcomes demonstrates the utility of \textit{platform portability}, as the same study design was able to be run both in-lab and on-road. This was made possible by the Portobello system's common model and vehicle localization; using robotics mapping and localization technology, we were able to capture surrounding environments for study and event staging for our on-road simulator. We anticipate that Portobello will advance the state of open-source and accessible driving simulation by extending the reach and translational capabilities of VR and XR driving simulation systems and thus, in turn, enable wider-scale development and testing of safe driving systems.

\section*{Open Science}
The source code for Portobello has been made publicly available. It can be accessed via the following link: \url{https://github.com/FAR-Lab/Portobello.git}.

\begin{acks}
% We thank all study participants.
This research was made possible by sponsorship by Toyota Research Institute and Woven by Toyota. Studies were conducted under Cornell Tech’s IRB Protocol \#IRB0008105. This work was also supported by a fellowship of the German Academic Exchange Service (DAAD).

\end{acks}

\balance
\bibliographystyle{ACM-Reference-Format}
\bibliography{references}
\end{document}